 \documentstyle[preprint,prl,tighten,aps,epsfig]{revtex}

\begin{document}

\draft

\preprint{\begin{tabular}{l}
\hbox to\hsize{\mbox{ }\hfill KIAS--P99110}\\%[-3mm] 
\hbox to\hsize{\mbox{ }\hfill hep--ph/9912330}\\%[-3mm] 
\hbox to\hsize{\mbox{ }\hfill December 1999}\\%[-3mm]
          \end{tabular}}

\title{Probing MSSM Higgs Sector with Explicit CP Violation at 
       a Photon Linear Collider}

\author{S.Y. Choi and Jae Sik Lee}
\address{Korea Institute for Advanced Study, Seoul 130--012, Korea}

\maketitle

\begin{abstract}
The $CP$ properties of Higgs bosons can be probed through their $s$--channel 
resonance productions via photon--photon 
collisions by use of circularly and/or linearly polarized backscattered
laser photons at a TeV--scale linear $e^+e^-$ collider.
Exploiting this powerful tool, we investigate in detail the Higgs sector 
of the minimal supersymmetric Standard Model with explicit $CP$ violation.
\end{abstract}


%\begin{multicols}{2}
%\narrowtext

%%%%%%%%%%%%%%%%%%%%%%%%%%%%%%
\section{Introduction}
\label{sec:introduction}
%%%%%%%%%%%%%%%%%%%%%%%%%%%%%%

% The experimental observation of Higgs boson(s) and the detailed confirmation 
% of their fundamental properties are crucial for our understanding of the 
% mechanism responsible for electroweak symmetry breaking. 

The minimal supersymmetric standard model (MSSM), of which the
Higgs sector is a well--defined two--Higgs--doublet model (2HDM), contains 
several $CP$--violating phases absent in the Standard Model (SM). 
In particular, the $CP$--violating phases of the higgsino mass parameter $\mu$ 
and of the stop and sbottom trilinear couplings $A_t$ and $A_b$ can affect 
the neutral Higgs sector significantly at the loop level due to the large 
Yukawa couplings, leading to a large mixing between the $CP$--even and 
$CP$--odd neutral Higgs bosons as well as to an induced relative phase 
$\xi$ between the vacuum expectation values of the two Higgs doublets 
\cite{DEMIR,PW}. 

The $CP$--violating phases do not have to be suppressed in order to satisfy 
the present experimental constraints from the electron and neutron electric 
dipole moments (EDMs) \cite{IN}. This has been shown, for example, in the 
context of the so--called effective supersymmetry (SUSY) model \cite{KAPLAN} 
where
the first and second generation sfermions are decoupled, but the third 
generation sfermions remain relatively light to preserve naturalness.  Based on
the scenarios of this type, the sensitivities of a variety of experimental
observables to $CP$ violation in the sfermion sector as well
as the neutral Higgs sector have been
recently examined in $B$ decays \cite{Many1}, supersymmetric dark--matter
searches, and  production of sparticles and Higgs bosons at LEP II, LHC and 
muon colliders \cite{DEMIR,Many2,Many3}. 

One of the cleanest determinations of the neutral Higgs sector $CP$ violation 
in the MSSM can be achieved by observing the $CP$ properties of all three 
neutral Higgs particles directly. In this light, the $s$--channel resonance 
production of neutral Higgs bosons in $\gamma\gamma$ collisions 
\cite{TWOPHOTON} has long been recognized as an important instrument to study 
the $CP$ properties of Higgs particles \cite{GG,LINEAR} at a linear $e^+e^-$ 
collider (LC) by use of polarized high energy laser lights obtained by 
Compton back--scattering of polarized laser light off the electron and positron
beams \cite{GKPST}. In the context of a general 2HDM involving a lot of
free parameters, the powerfulness of the production mechanism has been 
demonstrated by Grzadkowki and Gunion \cite{GG}.  On the contrary, the 
parameters determining the MSSM Higgs sector $CP$ violation are well 
defined so that the dependence of the $CP$ violation in the Higgs sector 
on the relevant
parameters can be explicitly studied.  So, in this paper, we demonstrate
that polarized back--scattered laser photons at a TeV--scale LC enable us to 
investigate the $CP$ violation of 
the Higgs sector in the MSSM through $s$--channel Higgs--boson production 
via $\gamma\gamma$ collisions in detail including its dependence on the
relevant SUSY parameters.

In Sec.~II we briefly review the $CP$--violating Higgs--boson mixing in the
MSSM induced at the loop-level from the stop and sbottom sectors due to the 
complex higgsino mass parameter, $\mu$, and trilinear couplings, $A_t$ and 
$A_b$.  Sec.~III is devoted to a model--independent description of the 
$s$--channel Higgs--boson production in polarized $\gamma\gamma$ collisions
leading to three polarization asymmetries expressed in terms of two
complex form factors for each neutral Higgs boson; one form factor is 
$CP$--even and the other one $CP$--odd. Then we briefly describe the 
mechanism of generating polarized back--scattered laser photons and
controlling the polarizations of the generated photons at a LC.  
In Sec.~IV, we present the analytic form of the $CP$--even and $CP$--odd
form factors explicitly in terms of the relevant SUSY parameters in the MSSM 
with explicit $CP$ violation. And in Sec.~V we investigate in detail
the dependence of the total production rates and the three polarization 
asymmetries on the $CP$--violating phases with the values of 
the other SUSY parameters fixed.
The conclusion is given in Sec.~VI and all the interaction Lagrangian
terms needed for the present work are listed in the Appendix.

%%%%%%%%%%%%%%%%%%%%%%%%%%%%%%%%%%%%%%%%%%%%%%%%%%%%%%%%
\section{CP--violating neutral Higgs--boson mixing}
\label{sec:mixing}
%%%%%%%%%%%%%%%%%%%%%%%%%%%%%%%%%%%%%%%%%%%%%%%%%%%%%%%%

The most general $CP$--violating Higgs potential of the
MSSM can conveniently be described by the effective Lagrangian \cite{PW}:
\begin{eqnarray}
{\cal L}_V&=&\mu^2_1(\Phi^\dagger_1\Phi_1) +\mu^2_1(\Phi^\dagger_2\Phi_2)
           +m^2_{12}(\Phi^\dagger_1\Phi_2)+m^{*2}_{12}(\Phi^\dagger_2\Phi_1)
	    \nonumber\\
         &&+\lambda_1(\Phi^\dagger_1\Phi_1)^2+\lambda_2(\Phi^\dagger_2\Phi_2)^2
           +\lambda_3(\Phi^\dagger_1\Phi_1)(\Phi^\dagger_2\Phi_2)
           +\lambda_4(\Phi^\dagger_1\Phi_2)(\Phi^\dagger_2\Phi_1)
            \nonumber\\
         &&+\lambda_5(\Phi^\dagger_1\Phi_2)^2
	   +\lambda^*_5(\Phi^\dagger_2\Phi_1)^2
           +\lambda_6(\Phi^\dagger_1\Phi_1)(\Phi^\dagger_1\Phi_2)
           +\lambda^*_6(\Phi^\dagger_1\Phi_1)(\Phi^\dagger_2\Phi_1)
            \nonumber\\
         &&+\lambda_7(\Phi^\dagger_2\Phi_2)(\Phi^\dagger_1\Phi_2)
           +\lambda^*_7(\Phi^\dagger_2\Phi_2)(\Phi^\dagger_2\Phi_1)\,,
\end{eqnarray}
where for convenience $\Phi_1=+i\tau_2\,H^*_1$ and $\Phi_2=H_2$ are
introduced instead of the conventional MSSM Higgs doublets $H_1$ and $H_2$.
At the tree level, $\mu^2_1=-m^2_1-|\mu|^2$ and $\mu^2_2=-m^2_2-|\mu|^2$
with $m^2_1$, $m^2_2$, and $m^2_{12}$ the soft--SUSY--breaking parameters
related to the Higgs sector, and the first four quartic couplings are 
determined solely by the SM gauge couplings; $\lambda_1=\lambda_2=
-\frac{1}{8}(g^2+g^{\prime 2})$, $\lambda_3 = -\frac{1}{4}(g^2-g^{\prime 2})$, 
and $\lambda_4=\frac{1}{2}g^2$, while the remaining three quartic couplings 
vanish; $\lambda_5=\lambda_6=\lambda_7=0$. 
Beyond the Born approximation, however, the quartic couplings $\{\lambda_5,
\lambda_6, \lambda_7\}$ can receive significant radiative corrections due to
large Yukawa couplings of the Higgs fields to the top/bottom and stop/sbottom
sectors. The analytic expressions of these parameters, which are in general 
complex, can be found in the Appendix of Ref.~\cite{PW}.

The $CP$--violating radiatively--corrected quartic couplings cause three 
physical neutral Higgs bosons to mix with one another. In order to describe the 
$CP$--violating Higgs--boson mixing, it is first of all necessary to determine 
the ground state of the Higgs potential.
To this end we introduce the linear decompositions of the Higgs fields
\begin{eqnarray}
\Phi_1=\left(\begin{array}{cc}
             \phi^+_1 \\
             \frac{1}{\sqrt{2}}(v_1+\phi_1+ia_1)
             \end{array}\right)\,, \qquad
\Phi_2={\rm e}^{i\xi}
       \left(\begin{array}{cc}
             \phi^+_2 \\
             \frac{1}{\sqrt{2}}(v_2+\phi_2+ia_2)
\end{array}\right)\,,
\end{eqnarray}
with $v_1$ and $v_2$ the moduli of the vacuum expectation values (VEVs) of the
Higgs doublets and $\xi$ is their relative phase. These VEVs and the relative 
phase can be determined by the minimization conditions on ${\cal L}_V$.
It is always guaranteed that one combination of the $CP$--odd  Higgs fields
$a_1$ and $a_2$ ($G^0=\cos\beta a_1+\sin\beta a_2$) defines a flat direction
in the Higgs potential and it is absorbed as the longitudinal component of the
$Z$ field. 
Denoting the remaining $CP$--odd state $a=-\sin\beta a_1
+\cos\beta a_2$, the neutral Higgs--boson mass matrix describing the 
mixing between $CP$--even and $CP$--odd fields 
in the $(a,\phi_1,\phi_2)$ basis is given by
\begin{eqnarray}
{\cal M}^2_N=\left(\begin{array}{cc}
{\cal M}^2_P     &   ({\cal M}^2_{SP})^T \\
{\cal M}^2_{SP}  &   {\cal M}^2_S     
\end{array}\right)\,.
\end{eqnarray}
The analytic form of the sub-matrices is given by
\begin{eqnarray}
&&{\cal M}^2_P=m^2_a=\frac{1}{s_\beta c_\beta}\left\{
              {\cal R}(m^2_{12}{\rm e}^{i\xi})
             +v^2\bigg[2{\cal R}(\lambda_5{\rm e}^{2i\xi})s_\beta c_\beta
             +\frac{1}{2}{\cal R}(\lambda_6{\rm e}^{i\xi})c^2_\beta
             +\frac{1}{2}{\cal R}(\lambda_7{\rm e}^{i\xi})s^2_\beta\bigg]
              \right\}\,, \nonumber\\
&&{\cal M}^2_{SP}=v^2\left(\begin{array}{c}
              {\cal I}(\lambda_5{\rm e}^{2i\xi})s_\beta
             +{\cal I}(\lambda_6{\rm e}^{i\xi})c_\beta\\
              {\cal I}(\lambda_5{\rm e}^{2i\xi})c_\beta
             +{\cal I}(\lambda_7{\rm e}^{i\xi})s_\beta
              \end{array}\right)\,, \nonumber\\
&&{\cal M}^2_S=m^2_a\left(\begin{array}{cc}
             s^2_\beta       & -s_\beta c_\beta\\
            -s_\beta c_\beta & c^2_\beta
              \end{array}\right)\nonumber\\
&& -v^2\left(\begin{array}{cc}
   2\lambda_1 c^2_\beta+2{\cal R}(\lambda_5{\rm e}^{2i\xi})s^2_\beta
  +2{\cal R}(\lambda_6{\rm e}^{i\xi})s_\beta c_\beta &
   \lambda_{34}s_\beta c_\beta+{\cal R}(\lambda_6{\rm e}^{i\xi})c^2_\beta
  +{\cal R}(\lambda_7{\rm e}^{i\xi})s^2_\beta \\  
   \lambda_{34}s_\beta c_\beta+{\cal R}(\lambda_6{\rm e}^{i\xi})c^2_\beta
  +{\cal R}(\lambda_7{\rm e}^{i\xi})s^2_\beta &
  2\lambda_2 s^2_\beta+2{\cal R}(\lambda_5{\rm e}^{2i\xi})c^2_\beta
  +2{\cal R}(\lambda_7{\rm e}^{i\xi})s_\beta c_\beta  
   \end{array}\right)\,.
\end{eqnarray}
The $CP$--even and $CP$--odd Higgs--boson states mix unless all the imaginary
parts of the parameters $\lambda_5,\lambda_6,\lambda_7$ vanish\footnote{ If
all the imaginary parts of $\lambda_i$ ($i=5,6,7$) are zero,  
the induced phase $\xi$ vanish as well.} and the symmetric Higgs--boson 
mass matrix ${\cal M}^2_N$ can
be diagonalized by a $3\times 3$ orthogonal matrix $O$ relating the weak
eigenstates to the mass eigenstates as 
%$(a,\phi_1,\phi_2)^T=O\,(H_3,H_2,H_1)^T$ 
%
\begin{eqnarray}
(a,\phi_1,\phi_2)^T=O\,(H_3,H_2,H_1)^T\,,
\end{eqnarray}
%
% as $O^T{\cal M}^2_N O={\rm diag}(m^2_{H_3},m^2_{H_2},m^2_{H_1})$
with the mass ordering of $m_{H_1}\leq m_{H_2}\leq m_{H_3}$.

The characteristic size of the $CP$--violating neutral Higgs--boson mixing  
is determined  by the factor 
\begin{eqnarray}
\frac{1}{32\pi^2}\frac{Y_f^4\,|\mu||A_f|}{M^2_{\rm SUSY}}\sin\Phi_{A_f\mu}\,,
\label{eq:size}
\end{eqnarray}
where $Y_f$ is the Yukawa coupling of the fermion $f$, $\Phi_{A_f\mu}={\rm
Arg}(A_f\mu)+\xi$ for $f=t,b$, and $M_{\rm SUSY}$ is a typical
SUSY--breaking scale, of which the square might be taken to be the average 
of the two sfermion masses squared, i.e.
$M_{\rm SUSY}^2=(m_{\tilde{f}_1}^2+m_{\tilde{f}_2}^2)/2$. 
The neutral Higgs--boson mixing modifies not only the Higgs mass spectra
but also their couplings to fermions, to sfermions, to the $W$ and $Z$ 
gauge bosons, and to the Higgs bosons themselves significantly.
(See the Appendix to find the interaction Lagrangian terms for the
couplings of the Higgs bosons with the fermions and bosons).
Therefore, the $CP$--violating mixing can affect the cross section of 
the process $\gamma\gamma\rightarrow H_i$ ($i=1,2,3$) significantly 
because the two--photon fusion process is mediated by loops of all charged 
particles with non--zero mass.

Although the stop and sbottom trilinear parameters are in general independent, 
we assume for our numerical analysis a universal trilinear parameter 
$A\equiv A_t=A_b$ 
and we vary the phase $\Phi=\Phi_\mu+\Phi_A$, where $\Phi_A$ is the phase
of the universal parameter $A$, over the range 
$[0^{\rm o},360^{\rm o}]$ and the charged Higgs--boson mass
$m_{H^{\pm}}$ up to 1 TeV. In addition, noting that the $CP$--violating 
phases could weaken the present experimental bounds on the lightest 
Higgs mass up to about 60 GeV \cite{PW}, we simply apply the lower mass
limit $M_{H_1}\geq 70$ GeV to the lightest Higgs--boson determining    
the lowest allowed value of $M_{H^\pm}$ for a given set of SUSY parameters. 
For the parameter $\tan\beta$, we take $\tan\beta=3$ or $10$ as the values 
representing the small and large $\tan\beta$ cases, respectively.
Finally, we take for the remaining dimensionful parameters 
\begin{eqnarray}
|A| = 0.4~{\rm TeV}\,, \  \ |\mu|=1.2~{\rm TeV}\,,\  \
M_{\rm SUSY}= 0.5~{\rm TeV}\,, \  \ \Delta_t=\Delta_b=M_{\rm SUSY}^2\,,
\label{eq:para}
\end{eqnarray}
with $\Delta_f= m_{\tilde{f}_2}^2-m_{\tilde{f}_1}^2$ for $f=t,b$,
safely avoiding the two--loop EDM constraints \cite{CKP}. The parameter
set (\ref{eq:para}) gives a rather large $CP$--violating neutral Higgs--boson
mixing as can be seen in Eq.~(\ref{eq:size}). However, the dependence of
our results on a different parameter set can be easily worked out.

%%%%%%%%%%%%%%%%%%%%%%%%%%%%%%%%%%%%%%%%%%%%%%%%%%%%%%%%%%%%%%%%%
\section{Two--photon fusion into Higgs bosons}
%%%%%%%%%%%%%%%%%%%%%%%%%%%%%%%%%%%%%%%%%%%%%%%%%%%%%%%%%%%%%%%%%

\subsection{Model independent description}

In the presence of the $CP$--violating neutral Higgs--boson mixing, 
the production amplitude
of the two--photon fusion process $\gamma\gamma\rightarrow H_i$ ($i=1,2,3$)
can be parameterized in a model--independent way in terms of two (complex) form 
factors $A_i$ and $B_i$ as
\begin{eqnarray}
{\cal M}(\gamma\gamma\rightarrow H_i)=M_{H_i}\frac{\alpha}{4\pi}
    \left\{A_i(s)\left[\epsilon_1\cdot\epsilon_2
                      -\frac{2}{s}(\epsilon_1\cdot k_2)(\epsilon_2\cdot k_1)
		 \right]
          -B_i(s)\frac{2}{s} \left\langle \epsilon_1\epsilon_2 k_1 k_2
                             \right \rangle \right\},
\label{me}
\end{eqnarray}
where $s$ is the c.m. energy squared of two colliding photons and the
notation $\langle\epsilon_1\epsilon_2 k_1 k_2\rangle$ stands for
the Lorentz invariant contraction $\epsilon_{\mu\nu\alpha\beta}\epsilon_1^\mu
\epsilon_2^\nu k_1^\alpha k_2^\beta $. 
In the two--photon c.m coordinate system with one photon momentum $\vec{k}_1$
along the positive $z$ direction and the other one $\vec{k}_2$ along the 
negative $z$ direction, the wave vectors $\epsilon_{1,2}$ of two photons 
are given by
\begin{eqnarray}
\epsilon_1(\lambda)=\epsilon^*_2(\lambda)
                   =\frac{1}{\sqrt{2}}\left(0,-\lambda,-i,0\right)\,.
\end{eqnarray}
where $\lambda=\pm 1$ denote the right and left photon helicities,
respectively.
Inserting the wave vectors into Eq.~(\ref{me}) we obtain the production
helicity amplitude for the photon fusion process as follows 
\begin{eqnarray}
{\cal M}_{\lambda_1\lambda_2}=-M_{H_i}\frac{\alpha}{4\pi}
     \left\{ A_i(s)\,\delta_{\lambda_1\lambda_2}
          +i \lambda_1 B_i(s)\delta_{\lambda_1\lambda_2}\right\}\,, 
\label{hamp}
\end{eqnarray}
with $\lambda_{1,2}=\pm$, yielding the absolute polarized amplitude squared 
\begin{eqnarray}
\overline{\left|{\cal M}\right|^2}=
           \overline{\left|{\cal M}\right|^2_0}\,\,
  \bigg\{             [1+\zeta_2\tilde{\zeta}_2]
      +{\cal A}_1\left[\zeta_2+\tilde{\zeta}_2\right]
      +{\cal A}_2\left[\zeta_1\tilde{\zeta}_3+\zeta_3\tilde{\zeta}_1\right]
      -{\cal A}_3\left[\zeta_1\tilde{\zeta}_1-\zeta_3\tilde{\zeta}_3\right]
                    \bigg\}\,,
\label{mesq}
\end{eqnarray}
with the Stokes parameters $\{\zeta_i\}$ and $\{\tilde{\zeta}_i\}$ ($i=1,2,3$)
of two photon beams, respectively.
The first factor in Eq.~(\ref{mesq}) is the unpolarized amplitude squared; 
\begin{eqnarray}
\overline{\left|{\cal M}\right|^2_0} =\frac{1}{4}
         \left\{\left|{\cal M}_{++}\right|^2
              +\left|{\cal M}_{--}\right|^2\right\}\,.
\end{eqnarray}
and three polarization asymmetries ${\cal A}_i$ ($i=1,2,3$) are defined in
terms of the helicity amplitudes and expressed in terms of the form factors
$A_i$ and $B_i$ as
\begin{eqnarray}
{\cal A}_1&=&
   \frac{\left|{\cal M}_{++}\right|^2-\left|{\cal M}_{--}\right|^2}
        {\left|{\cal M}_{++}\right|^2+\left|{\cal M}_{--}\right|^2}
    =\frac{2{\cal I}(A(s)B(s)^*)}{\left|A(s)\right|^2+\left|B(s)\right|^2}\,, 
     \nonumber \\
{\cal A}_2&=&
   \frac{2{\cal I}({\cal M}_{--}^*{\cal M}_{++})}
        {\left|{\cal M}_{++}\right|^2+\left|{\cal M}_{--}\right|^2}
    =\frac{2{\cal R}(A(s)B(s)^*)}{\left|A(s)\right|^2+\left|B(s)\right|^2}\,, 
     \nonumber \\
{\cal A}_3&=&
   \frac{2{\cal R}({\cal M}_{--}^*{\cal M}_{++})}
        {\left|{\cal M}_{++}\right|^2+\left|{\cal M}_{--}\right|^2}
    =\frac{\left|A(s)\right|^2-\left|B(s)\right|^2}
          {\left|A(s)\right|^2+\left|B(s)\right|^2}\,. 
\end{eqnarray}
In the $CP$--invariant theories with real couplings, the form factors
$A_i$ and $B_i$ cannot be simultaneously non--vanishing so that 
they should satisfy the relations;
${\cal A}_1={\cal A}_2=0$ and ${\cal A}_3=+1(-1)$ depending on whether the
Higgs boson is a pure $CP$--even ($CP$--odd) state. In other words,
${\cal A}_1\neq 0$, ${\cal A}_2\neq 0$ and/or 
$\left|{\cal A}_3\right|<1$ ensure a simultaneous existence of non-zero 
$A_i$ and $B_i$ implying $CP$ violation. Note that the asymmetry ${\cal A}_1$ 
is non--vanishing only when $A_i$ and $B_i$ have a finite relative phase. 
As will be seen explicitly in the next section, even for the real MSSM
couplings $A_i(s)$ and $B_i(s)$ could be complex because of the 
nontrivial developments of the imaginary parts for the Higgs masses 
larger than twice the loop masses. 

In the narrow--width approximation, the partonic cross section of the
$s$-channel Higgs--boson production $\gamma\gamma\rightarrow H_i$ can be 
expressed as 
\begin{eqnarray}
\sigma(\gamma\gamma \rightarrow H_i)
% &=&\frac{8\pi^2}{M_{H_i}}\Gamma(H_i\rightarrow\gamma\gamma)
% \frac{M\Gamma_{H_i}/\pi}
% {(s_{\gamma\gamma}-M_{H_i}^2)^2+(M_{H_i}\Gamma_{H_i})^2}
% \nonumber \\
= \frac{\pi}{M^4_{H_i}}\,\overline{\left|{\cal M}\right|^2_0}\,
  \delta\left(1-\frac{M_{H_i}^2}{s}\right)\,
  \equiv \hat\sigma_0(H_i)\, \delta\left(1-\frac{M_{H_i}^2}{s}\right)\,,
\label{eq:partonic}
\end{eqnarray}
which is eventually to be 
folded with a realistic photon luminosity spectrum. Certainly, it is 
recommended to use as hard photon spectra as possible and to have the 
capability of controlling photon polarizations with ease. 
Such a high energy polarized photon beam is available using Compton 
backscattering of laser light off the electron or positron beams. 
Although the detailed description of 
the generation mechanism has been provided in literature, we will 
describe for our purpose the mechanism briefly in the following subsection.

\subsection{Polarized high energy laser back--scattered photons}

High energy colliding beams of polarized photons can be generated by
Compton backscattering of polarized laser light on (polarized) 
electron/positron bunches of $e^+e^-$ linear colliders\footnote{In the present
work the electron and positron beams are assumed to be unpolarized.
It is however straightforward to take into account polarized electron 
and positron beams.}.
The polarization transfer from the laser light to the high energy photons is
described by three Stokes parameters $\zeta_{1,2,3}$; $\zeta_2$ is the
degree of circular polarization and $\{\zeta_3,\zeta_1\}$ the degree of
linear polarization transverse and normal to the plane defined by the
electron direction and the direction of the maximal linear polarization
of the initial laser light. Explicitly, the Stokes parameters take the 
form \cite{GKPST}:
\begin{eqnarray}
\zeta_1= \frac{f_3(y)}{f_0(y)}\,P_t\,\sin{2\kappa}\,,\qquad
\zeta_2=-\frac{f_2(y)}{f_0(y)}\,P_c\,,\qquad
\zeta_3= \frac{f_3(y)}{f_0(y)}\,P_t\,\cos{2\kappa}\,,
\label{stokes}
\end{eqnarray}
where $y$ is the energy fraction of the back--scattered photon with respect to
the initial electron energy $E_e$, $\{P_c, P_t\}$ are the degrees of 
circular and transverse polarization of the initial laser light, and
$\kappa$ is the azimuthal angle between the directions of initial photon 
and its maximum linear polarization.  Similar relations can be obtained 
for the Stokes parameters $\tilde{\zeta}$ of the opposite high energy photons
by replacing $(P_c, P_t, \kappa)$ with $(\tilde{P}_c, \tilde{P}_t, 
-\tilde{\kappa})$.  The functions $f_0$, $f_2$, and $f_3$ determining the
photon energy spectrum and the Stokes parameters are given by
\begin{eqnarray}
f_0(y)=\frac{1}{1-y}+1-y-4r(1-r) \,, \ \ 
f_2(y)=(2r-1)\left(\frac{1}{1-y}+1-y \right) \,, \ \
f_3(y)=2r^2 \,, 
\end{eqnarray}
with $r=y/x(1-y)$ and $x=4E_e\omega_0/m_e^2\approx 15.4(E_e[{\rm TeV}])
(\omega_0[{\rm eV}])$ for the initial laser energy $\omega_0$.
We observe in Eqs.~(\ref{mesq}) and (\ref{stokes}) that the linear 
polarization of the high energy photon beam is proportional to $P_t$ whereas 
the circular polarization is proportional to $P_c$. Therefore, it is
necessary to have both circularly and linearly polarized photons in
order to measure all the polarization asymmetries ${\cal A}_{1,2,3}$ and
as a result the complex form factors $A_i(s)$ and $B_i(s)$.  

After folding the luminosity spectra of two photon beams, the event rate of 
the Higgs boson production via two--photon fusion is given by
\begin{eqnarray}
\frac{{\rm d}N}{{\rm d}\eta}&=&\frac{{\rm d}L_{\gamma\gamma}}{{\rm d}\eta}
\frac{M^4_{H_i}}{\pi}\,\hat{\sigma}_0
 \left\{1+P_c\tilde{P}_c\, \frac{\langle f_2 * f_2 \rangle_\eta}{\langle 
                                         f_0 * f_0 \rangle_\eta} \right.
-{\cal A}_1 \left(P_c+\tilde{P}_c\right)
 \frac{\langle f_0 * f_2 \rangle_\eta}{\langle 
               f_0 * f_0 \rangle_\eta}\nonumber \\
&&{ }\hskip 2cm +P_t\tilde{P}_t\bigg[{\cal A}_2\sin2(\kappa-\tilde{\kappa})+
   {\cal A}_3\cos2(\kappa-\tilde{\kappa})\bigg]
   \left.\frac{\langle f_3 * f_3 \rangle_\eta}{\langle 
                       f_0 * f_0 \rangle_\eta}\right\} \,,
\label{eq:fold}
\end{eqnarray}
where $\frac{{\rm d}L_{\gamma\gamma}}{{\rm d}\eta}$ is the two---photon 
luminosity function depending on the details such as the $e$-$\gamma$ 
conversion factor and the shape of the electron/positron bunches
\cite{GKPST}, and $\eta\equiv s/s_{ee}=m^2_{H_i}/s_{ee}$.
The luminosity correlation functions $\langle f_i * f_j \rangle_\eta$ 
are defined as
\begin{eqnarray}
\langle f_i * f_j \rangle_\eta=\frac{1}{N^2}
       \int^{y_{\rm max}}_{\eta/y_{\rm max}}\frac{{\rm d}y}{y}f_i(y)f_j(\eta/y)\,,
\end{eqnarray}
with the normalization $N=\int^{y_{\rm max}}_0{\rm d}y\,f_0(y)$ and 
$y_{\rm max}=x/(1+x)$.

We show in Fig.~1 the correlation function $\langle f_0 * f_0 \rangle_\eta$ 
and the three ratios of the correlation functions appearing in 
Eq.~(\ref{eq:fold}) for $x=0.5$ (solid line), $1.0$ (dashed line), and 
$4.83$ (dotted line).
For a larger value of $x$, the correlation function $\langle f_0 * f_0 
\rangle_\eta$, to which ${\rm d}L_{\gamma\gamma}/{\rm d}\eta$ is proportional,
becomes more flat and the maximal obtainable photon energy fraction becomes 
closer to the electron beam energy. Exploiting this feature appropriately
could facilitate Higgs--boson searches at a photon linear collider.
The three figures for the ratios of correlation functions clearly show
that the maximal sensitivity to each polarization asymmetry 
${\cal A}_i$ can be acquired near the the maximal value of  
$\eta=y_{\rm max}^2$. Therefore, once the Higgs--boson masses are
known, one can obtain the maximal sensitivities by tuning the initial electron 
energy to be 
\begin{eqnarray}
E_e=\left(\frac{1+x}{2x}\right) M_{H_i} \,.
\end{eqnarray}
On the other hand, the ratio $\langle f_3 * f_3 \rangle_\eta/\langle f_0 * f_0
\rangle_\eta$ is larger for a smaller value of $x$ and for a given $x$
the maximum value of the ratio is given by
\begin{eqnarray}
\left(\frac{\langle f_3*f_3 \rangle_\eta}{\langle 
                    f_0*f_0 \rangle_\eta}\right)_{\rm max}
=\left[\frac{2(1+x)}{1+(1+x)^2}\right]^2\,.
\end{eqnarray}
Consequently, it is necessary to take a small $x$ and a high $E_e$ 
by changing the laser beam energy $\omega_0$ so as to acquire the highest
sensitivity to $CP$ violation in the neutral Higgs sector.

%%%%%%%%%%%%%%%%%%%%%%%%%%%%%%%%%%%%%%%%%%%%%%%
\section{The MSSM with explicit CP violation}
%%%%%%%%%%%%%%%%%%%%%%%%%%%%%%%%%%%%%%%%%%%%%%%

In this section, we consider the MSSM as a specific 2HDM, in which the Higgs
sector $CP$ violation is induced at the loop level from the stop and sbottom
sectors defined by the trilinear parameters $A_{t,b}$ and the higgsino
mass parameter $\mu$ \cite{PW}. Then, we derive the explicit form of the 
form factors $A_i(s)$ and $B_i(s)$ by calculating the loop contributions from 
the bottom and top quarks, the charged Higgs boson, the $W$ boson 
as well as the lighter top and bottom squarks\footnote{There might exist the
contributions from other charged particles such as charginos and heavier
stop and sbottoms. The contributions from the charginos are neglected in the
present work by assuming them to be heavy while those from the heavier
sfermions are neglected because they are very heavy for the parameter set
(\ref{eq:para}).}.

Taking the sum of all the charged particle contributions, 
we obtain for the form factor $A_i$ at $s=M^2_{H_i}$: 
\begin{eqnarray}
A_i\left(s=M_{H_i}^2\right)=\sum_{f=t,b}A_i^f
                  +\sum_{\tilde{f}_j=\tilde{t}_1,\tilde{b}_1}A_i^{\tilde{f}_j}
                  +A_i^{H^{\pm}} +A_i^{W^{\pm}} \,,
\end{eqnarray}
with the $CP$-even functions  
\begin{eqnarray}
&& A_i^f=-2\left(\sqrt{2}G_F\right)^{1/2} M_{H_i} N_c e_f^2
          \left(\frac{v_f^i}{R_\beta^f}\right) F_{sf}(\tau_{if})\,,\nonumber \\
&& A_i^{\tilde{f}_j}=\frac{M_{H_i}N_c\, e_f^2\, g^i_{\tilde{f}_j\tilde{f}_j}}
    {2 m_{\tilde{f}_j}^2}F_{0}(\tau_{i\tilde{f}}) \,,\nonumber\\
&& A_i^{H^{\pm}}=\frac{M_{H_i}\,v\,C_i}{2 m_{H^{\pm}}^2}F_{0}(\tau_{iH})\,, 
   \nonumber \\
&& A_i^{W^{\pm}}=\left(\sqrt{2}G_F\right)^{1/2} M_{H_i}
   (c_\beta O_{2,4-i}+s_\beta O_{3,4-i}) F_{1}(\tau_{iW}) \,,
\end{eqnarray}
where $N_c=3$, the scaling variable $\tau_{ix}=m^2_{H_i}/4m^2_x$ for 
$m_x=m_f,m_{\tilde{f}_i},M_{H^\pm}, m_{W^\pm}$, and the definition of 
all the {\it real} couplings $v_f^i$, 
$g^i_{\tilde{t}_1\tilde{t}_1}$, $g^i_{\tilde{b}_1\tilde{b}_1}$, 
$C_i$, and $O_{\alpha,4-i}$ as well as $R_\beta^f$ is given in the Appendix.
On the other hand, the form factor $B_i$ related to the fermionic triangle
anomaly has the contributions only from the top and bottom quarks and it
takes the form  
\begin{eqnarray}
B_i\left(s=M_{H_i}^2\right)=-2\left(\sqrt{2}G_F\right)^{1/2}M_{H_i} N_c
  \sum_{f=t,b} e_f^2\left(\frac{\bar{R}^f_\beta a^i_f}{R^f_\beta} \right) 
   F_{pf}(\tau_{if}) \,,
\end{eqnarray}
where the definition of the real couplings $a^i_f$ also can be found in the
Appendix. The form factors $F_{sf}$, $F_{pf}$, $F_0$, and $F_1$ can be
expressed as
\begin{eqnarray}
F_{sf}(\tau)&=&\tau^{-1}\,[1+(1-\tau^{-1}) f(\tau)]\,,\qquad
F_{pf}(\tau)=\tau^{-1}\,f(\tau)\,,\nonumber\\
F_0(\tau)&=&\tau^{-1}\,[-1+\tau^{-1}f(\tau)]\,,\qquad\ \
F_1(\tau)=2+3\tau^{-1}+3\tau^{-1} (2-\tau^{-1} )f(\tau) \,,
\label{formfactor}
\end{eqnarray}
in terms of the scaling function $f(\tau)$ \cite {HHG}:
\begin{eqnarray}
f(\tau)=-\frac{1}{2}\int_0^1\frac{{\rm d}y}{y}\ln\left[1-4\tau y(1-y)\right]
       =\left\{\begin{array}{cl}
           {\rm arcsin}^2(\sqrt{\tau}) \,,   & \qquad \tau\leq 1\,, \\
   -\frac{1}{4}\left[\ln \left(\frac{\sqrt{\tau}+\sqrt{\tau-1}}{
                                     \sqrt{\tau}-\sqrt{\tau-1}}\right)
                    -i\pi\right]^2\,, & \qquad \tau\geq 1\,.
\end{array}\right.
\end{eqnarray}
It is clear that the imaginary parts of the form factors are 
developed for the Higgs-boson mass greater than twice the mass of 
the charged particle running in the loop, i.e. $\tau\geq 1$ as shown explicitly
in Fig.~2. In the presence of weak $CP$--violating phases, this development 
of the imaginary parts can lead to a nonzero polarization asymmetry 
${\cal A}_1$.

%%%%%%%%%%%%%%%%%%%%%%%%%%%%%%%
\section{Numerical Results}
%%%%%%%%%%%%%%%%%%%%%%%%%%%%%%%

Based on the parameter set (\ref{eq:para}) for two values of $\tan\beta=3,10$, 
we investigate in detail the dependence of the unpolarized part 
$\hat\sigma_0(H_i)$ and the three polarization asymmetries $A_i$ for
the production of all the three Higgs bosons via two--photon fusion.
However, we do not present the cross sections folded with 
the photon luminosity spectra explicitly because they can be obtained 
in a rather straightforward way from the partonic cross sections.
We do not take into account the QCD radiative corrections either, but 
for the details we refer to Ref.~\cite{SDGZ}.

Fig.~3 shows the unpolarized cross section $\hat{\sigma}_0(H_i)$ in units of
fb as a function of each Higgs--boson mass $M_{H_i}$ for five different
values of the $CP$--violating phase; $\Phi=0^{\rm o}$ (thick solid 
line), $40^{\rm o}$(solid line), $80^{\rm o}$(dashed line), 
$120^{\rm o}$(dotted line), and $160^{\rm o}$(dash--dotted line)
by taking $\tan\beta=3$ (left column) and $\tan\beta=10$ (right column),
respectively. We note that for $\tan\beta=3$ the phase $\Phi=160^{\rm o}$ is 
not allowed because it violates the bound $\Delta_f\leq
M^2_{\rm SUSY}$ required to ensure the validity of the loop corrections to the
Higgs potential \cite{GLUE,PW}. 
The charged Higgs--boson contributions to the production cross 
sections are negligible for every Higgs boson because the charged 
Higgs--boson--pair threshold is much higher than $M_{H_1}$ and nearly 
twice as large as the heavier Higgs--boson masses $M_{H_{2,3}}$ 
for both $\tan\beta=3$ and $\tan\beta=10$.

Firstly, we consider the production of the lightest Higgs boson $H_1$ 
with its mass $m_{H_1}$ less than 130 GeV. For $\tan\beta=3$ 
the cross section $\hat\sigma_0(H_1)$ is dominated by the $W$ 
loop contributions which are hardly dependent on the phase $\Phi$.
On the other hand, for $\tan\beta=10$ the $W$ loop contributions
become very sensitive to the $CP$--violating phase $\Phi$
and so does the cross section, especially for a small Higgs--boson mass. 
Numerically, the first frame
of the right column in Fig.~3 shows that the decrease of the cross section 
with increasing $\Phi$ is much more significant for a smaller Higgs boson 
mass.

Secondly, we consider the production of the heavier Higgs bosons $H_2$ and
$H_3$. In the case of $\tan\beta=3$, we note:
\begin{itemize}
\item In the $CP$--invariant theories, i.e. when $\Phi=0^{\rm o}$,
      the partonic cross section $\hat\sigma_0(H_2)$ has the contributions
      only from the top--quark loop\footnote{Certainly there exists the 
      the bottom--quark loop contribution. However, it is negligible 
      because of its very small Yukawa coupling for $\tan\beta=3$.}.
      This is evident by noting that the thick solid line has a single
      peak at the top--quark--pair threshold.
\item As the phase $\Phi$ increases, the contributions of the $W$ boson 
      and light stop loops become sizable; the former contributions are
      evident for smaller values of $M_{H_2}$ and the latter ones for
      larger values of $M_{H_2}$. This is because the Higgs boson 
      $H_2$ develops its $CP$--even component with non--trivial $\Phi$. 
      The stop contributions always increase the cross section 
      $\hat\sigma_0(H_2)$ above the 
      top--quark--pair threshold. It is noted that there exist small changes 
      below the top--quark--pair threshold due to the new $W$
      loop contributions.
\item In the $CP$--invariant case the heaviest Higgs boson $H_3$ is 
      $CP$--even so that the main contributions to  the cross 
      section $\hat\sigma_0(H_3)$ are from the top--quark loop
      with the function $F_{sf}(\tau)$ smoother than the other functions
      around the top--quark--pair threshold. The steep rise of the cross 
      section with decreasing $M_{H_3}$ is due to the $W$ loop 
      contribution near the $W$--pair threshold. Near the lighter stop--pair
      threshold one can see a small bump due to the lighter stop loop
      contribution.
\item The cross section $\hat\sigma_0(H_3)$ increases significantly with 
      the phase $\Phi$ due to the lighter stop--loop contributions except 
      for smaller values of $M_{H_3}$.
\end{itemize}
In the case of $\tan\beta=10$, we may draw the similar arguments as in
the case of $\tan\beta=3$ except for two significantly different aspects;
(i) the bottom as well as sbottom loop contributions become significant. 
For example, the small increase of the lower end tail of the thick solid 
line in the second figure of the right column for $\Phi=0^{\rm o}$ 
is caused by the bottom--quark--loop
contribution; (ii)  the mass ordering of the heavy $CP$--odd and
$CP$-even Higgs states is completely interchanged for large $\Phi$, that is 
to say, $H_2$ becomes $CP$--even and $H_3$ becomes $CP$--odd for large $\Phi$. 
(This aspect can be seen more clearly by the polarization asymmetry 
${\cal A}_3$ for $\tan\beta=10$ as will be shown in the following.)
This (almost) complete interchange of the $CP$ properties is reflected in the
similar patterns between $\hat\sigma_0(H_2, \Phi=0^{\rm o})$ and 
$\hat\sigma_0(H_3, \Phi=160^{\rm o})$ as well as between
$\hat\sigma_0(H_2, \Phi=160^{\rm o})$ and $\hat\sigma_0(H_3, \Phi=0^{\rm o})$
as can be checked in Fig.~3.

The cross sections are not genuine $CP$--odd observables so that they cannot
allow us to measure the $CP$--violating phase directly. On the contrary, as
mentioned before, the measurements of the polarization asymmetries enable
us to probe $CP$ violation directly. 
In order to discuss the polarization asymmetries more efficiently, we first 
consider the case of $\tan\beta=3$.
Fig.~4 shows the polarization asymmetries $A_i$ as a function of each
Higgs-boson mass for four values of the  $CP$--violating phase;
$\Phi=0^{\rm o}$ (thick solid line), $40^{\rm o}$ (solid line), $80^{\rm o}$
(dashed line), and $120^{\rm o}$(dotted line). The left column is for 
the polarization asymmetry ${\cal A}_1$, the middle column for the 
polarization asymmetry ${\cal A}_2$, and the right column for the
polarization asymmetry ${\cal A}_3$. The polarization asymmetries for
the $CP$--violating phase $\Phi$ larger than $180^{\rm o}$ can be read
by the relations 
\begin{eqnarray}
{\cal A}_{1,2}(\Phi)=-{\cal A}_{1,2}(360^{\rm o}-\Phi)\,,\qquad
{\cal A}_3(\Phi)    =+{\cal A}_3(360^{\rm o}-\Phi)\,,
\end{eqnarray}
reflecting the fact that the asymmetries ${\cal A}_{1,2}$ are $CP$--odd 
observables and the asymmetry ${\cal A}_3$ a $CP$--even observable.
We note from Fig.~4 that (i) for the lightest Higgs boson $H_1$ the asymmetry 
${\cal A}_1$ is smaller than 1\% and  the deviation of the asymmetry 
${\cal A}_3$ from the unity is negligible, whereas the asymmetry ${\cal A}_2$ 
can be as large as 5\% if the Higgs boson is light, (ii)  
for the heavier Higgs bosons $H_2$ and $H_3$, the asymmetries ${\cal A}_i$ 
are determined mainly by the top and stop loop contributions above 
the top--pair threshold, while they are determined by the top and $W$ 
loop contributions below the top--pair threshold, and (iii) there exists 
an additional small contribution from the bottom loop contribution to the 
asymmetry ${\cal A}_1$ below the $W$--pair threshold.

The same polarization asymmetries are displayed as a function of each 
Higgs-boson
mass for the same set of the phases in Fig.~5 by taking  a larger value of
$\tan\beta=10$ rather than $\tan\beta=3$ as in Fig.~4. 
Unlike the small $\tan\beta$ case, $\Phi=160^{\rm o}$ is allowed so that 
the polarization asymmetries (dash--dotted line)
for the phase value are also considered in the discussion.
We note from Fig.~5 that (i) even for the lightest Higgs boson $H_1$, 
the polarization asymmetries ${\cal A}_1$ and ${\cal A}_2$ can be as large 
as 40\% and 7\%, respectively and the deviation of ${\cal A}_3$ from the unity
can be as large as 10\% unlike the small $\tan\beta$ case, (ii)
the polarization asymmetries ${\cal A}_i$ for the heavier Higgs bosons 
$H_2$ and $H_3$ are determined mainly by the top/bottom and stop/sbottom 
loop contributions above the top-quark--pair threshold,
while below the top-quark--pair threshold the main contributions come from 
the top/bottom and $W$ loops,
and (iii) for large $\Phi$ values ($\Phi=120^{\rm o}$ and $160^{\rm o}$)
the $CP$--parities of the heavier Higgs bosons are interchanged as can be
checked in the two lower frames of the right column for ${\cal A}_3$. 

Combining the numerical results from Figs.~4 and 5, we can conclude that
for a small $\tan\beta$ the best $CP$ observable is the polarization asymmetry
${\cal A}_2$ whereas for a large $\tan\beta$ the polarization asymmetry
$A_1$ is the most powerful observable for detecting $CP$--violation in the
production of the lightest Higgs boson. Therefore, for a small $\tan\beta$
it is necessary to prepare the colliding photon beams with large linear 
polarizations, but those with large 
circular polarizations for a large $\tan\beta$.
On the other hand, all the polarization asymmetries ${\cal A}_i$ for 
the heavy Higgs bosons $H_2$ and $H_3$ are mostly very sensitive
to the the $CP$--violating phase $\Phi$ irrespective of the value of
$\tan\beta$. Moreover, the polarization asymmetries ${\cal A}_1$ and
${\cal A}_2$ are complementary in the sense that in the mass range where
one asymmetry is insensitive to $\Phi$ the other one is always sensitive
to the phase.

%%%%%%%%%%%%%%%%%%%%%%
\section{Conclusion}
%%%%%%%%%%%%%%%%%%%%%%

We have studied the possibility of measuring the $CP$ properties 
of the Higgs bosons in the MSSM with explicit CP violation in the
production of all the three neutral Higgs bosons in two--photon fusion
with polarized back--scattered laser photons. 

Our detailed analysis has clearly shown that collisions 
of polarized photons can provide a significant opportunity for detecting 
$CP$ violation in the MSSM Higgs sector induced at the
loop level from the stop and sbottom sectors.

\vskip 0.3cm

%%%%%%%%%%%%%%%%%%%%%%%%%%%%%%%%%%%%%%%%%%%%%%%%%%%%%%%%
\section*{Acknowledgments}
%%%%%%%%%%%%%%%%%%%%%%%%%%%%%%%%%%%%%%%%%%%%%%%%%%%%%%%%

The work of SYC was supported by the Korea Science and Engineering 
Foundation (KOSEF) through the KOSEF--DFG large collaboration project, 
Project No.~96--0702--01-01-2.

\setcounter{equation}{0}
\renewcommand{\theequation}{A\arabic{equation}}

%

%

%\newpage
\setcounter{equation}{0}
\renewcommand{\theequation}{A\arabic{equation}}

%%%%%%%%%%%%%%%%%%%%%%%%%%%%
\section*{Appendix}
%%%%%%%%%%%%%%%%%%%%%%%%%%%%

\def\beq{\begin{equation}}
\def\eeq{\end{equation}}
\def\beqar{\begin{eqnarray}}
\def\eeqar{\end{eqnarray}}
\def\Re#1{{\cal R}\left({#1}\right)}
\def\Im#1{{\cal I}\left({#1}\right)}
\def\sb#1{{s_\beta^{#1}}}
\def\cb#1{{c_\beta^{#1}}}
\def\tb#1{{t_\beta^{#1}}}
\def\l#1{\lambda_{#1}{\rm e}^{i\xi}}
\def\half{\frac{1}{2}}

In this Appendix, we present all the MSSM interaction Lagrangians needed to
calculate the contributions of the fermions, the charged Higgs boson,
the charged gauge bosons, and the charged sfermions to the two--photon 
fusion processes $\gamma\gamma\rightarrow H_i$ ($i=1,2,3$). 
They are classified in the following four categories; 

\begin{enumerate}

\item \underline{\bf Higgs--fermion--fermion couplings}:
\begin{eqnarray}
&& {\cal L}_{H\bar{f}f}=-\frac{g m_f}{2m_W}\, 
          \bar{f}\left[\left(\frac{v^i_f}{R^f_\beta}\right)
        -i\left(\frac{\bar{R}^f_\beta a^i_f}{R^f_\beta}\right)
                 \gamma_5\right]f\, H_i\,, \nonumber \\
&& R^f_\beta =\left\{\begin{array}{c} c_\beta \\ s_\beta\end{array}\right.\,,
\  \
\bar{R}^f_\beta =\left\{\begin{array}{c} s_\beta\\    
c_\beta\end{array}\right.\,,
\,\, 
v^i_f =\left\{\begin{array}{c} O_{2,4-i}\\ O_{3,4-i}    
\end{array}\right.\,,
\,\, 
a^i_f =\left\{\begin{array}{c} O_{1,4-i}\\ O_{1,4-i}    
\end{array}\right.
\,\,
\begin{array}{l} {\rm for}\ \ f=(l:d) \\ {\rm for}\ \ f=(u)     
\end{array}\,.
\end{eqnarray}
The $CP$--violating neutral Higgs--boson mixing induces a simultaneous 
coupling of $H_i$ ($i=1,2,3$) to $CP$--even and $CP$--odd fermion bilinears 
$\bar{f}f$ and $\bar{f} i\gamma_5 f$.

\item  \underline{\bf Higgs--{\it H}$^+$--{\it H}$^-$ vertices}:
\begin{eqnarray}
{\cal L}_{HH^+H^-}=v\, C_i H_i H^+ H^- \, ~~{\rm with}~~
C_i=\sum_{\alpha=1,2,3} O_{\alpha, 4-i} c_\alpha
\end{eqnarray}
with
\begin{eqnarray}
c_1 &=& 2 \sb{} \cb{} \Im{\lambda_5{\rm e}^{2i\xi}}
      -\sb{2} \Im{\l{6}}-\cb{2} \Im{\l{7}} \, , \nonumber \\ 
c_2 &=& 2 \sb{2}\cb{}\lambda_1+\cb{3}\lambda_3-\sb{2}\cb{}\lambda_4\nonumber\\
    && -2\sb{2}\cb{}\Re{\lambda_5{\rm e}^{2i\xi}}
      +\sb{}(\sb{2}-2\cb{2})\Re{\l{6}}+\sb{}\cb{2}\Re{\l{7}} \, ,\nonumber \\ 
c_3 &=& 2 \cb{2}\sb{}\lambda_2+\sb{3}\lambda_3-\cb{2}\sb{}\lambda_4\nonumber\\ 
    && -2\cb{2}\sb{}\Re{\lambda_5{\rm e}^{2i\xi}}
      +\cb{}\sb{2}\Re{\l{6}}+\cb{}(\cb{2}-2\sb{2})\Re{\l{7}}  \, . 
\end{eqnarray}
Note that the coupling $c_1$ vanishes in the CP--invariant theories 
where the imaginary parts of the quartic couplings $\lambda_{5,6,7}$ and
the induced phase are zero.

\item \underline{\bf Higgs--sfermion-sfermion vertices}:
\begin{eqnarray}
{\cal L}_{H_i\tilde{f}_j\tilde{f}_k}&=&g^i_{\tilde{f}_j\tilde{f}_k}
        \tilde{f}^*_j \tilde{f}_k H_i\,,\nonumber\\
g^i_{\tilde{f}_j\tilde{f}_k}&=& 
%C^f_{\alpha;\beta\gamma}
%O_{\alpha, 4-i} ( P_f\, U_f)^*_{\beta j} (P_f\,U_f)_{\gamma k}
%= 
\tilde{C}^f_{\alpha;\beta\gamma}
O_{\alpha, 4-i} (U_f)^*_{\beta j} (U_f)_{\gamma k}\,,
\end{eqnarray}
where the index $\alpha$ denotes three neutral Higgs--boson fields 
$\{a,\phi_1,\phi_2\}$ and $\{\beta,\gamma\}$ the chiralities $\{L,R\}$. 
The unitary matrix $U_f$ diagonalizes the usual sfermion mass matrix 
${\cal M}^2_{\tilde{f}}$ as $U^\dagger_f\,{\cal M}^2_{\tilde{f}}\, U_f
={\rm diag}(m^2_{\tilde{f}_1}, m^2_{\tilde{f}_2})$ with
$m_{\tilde{f}_1}\leq m_{\tilde{f}_2}$. 
For the details of the sfermion mixing and the explicit form of the
chiral couplings $\tilde{C}^f_{\alpha;\beta\gamma}$, we refer to
Ref.~\cite{GLUE}.

\item \underline{\bf Higgs--$W$--$W$ vertices}:
\begin{eqnarray}
{\cal L}_{HW^+W^-}=gm_W(c_\beta O_{2,4-i}+s_\beta O_{3,4-i})
    \, H_i W^+_\mu W^{-\mu}\,.
\end{eqnarray}
\end{enumerate}

    %%%%%%%%%%%%%%%%%%%% Figures %%%%%%%%%%%%%%%%%%%%%%%%%

\begin{figure}
\begin{center}
\hbox to\textwidth{\hss\epsfig{file=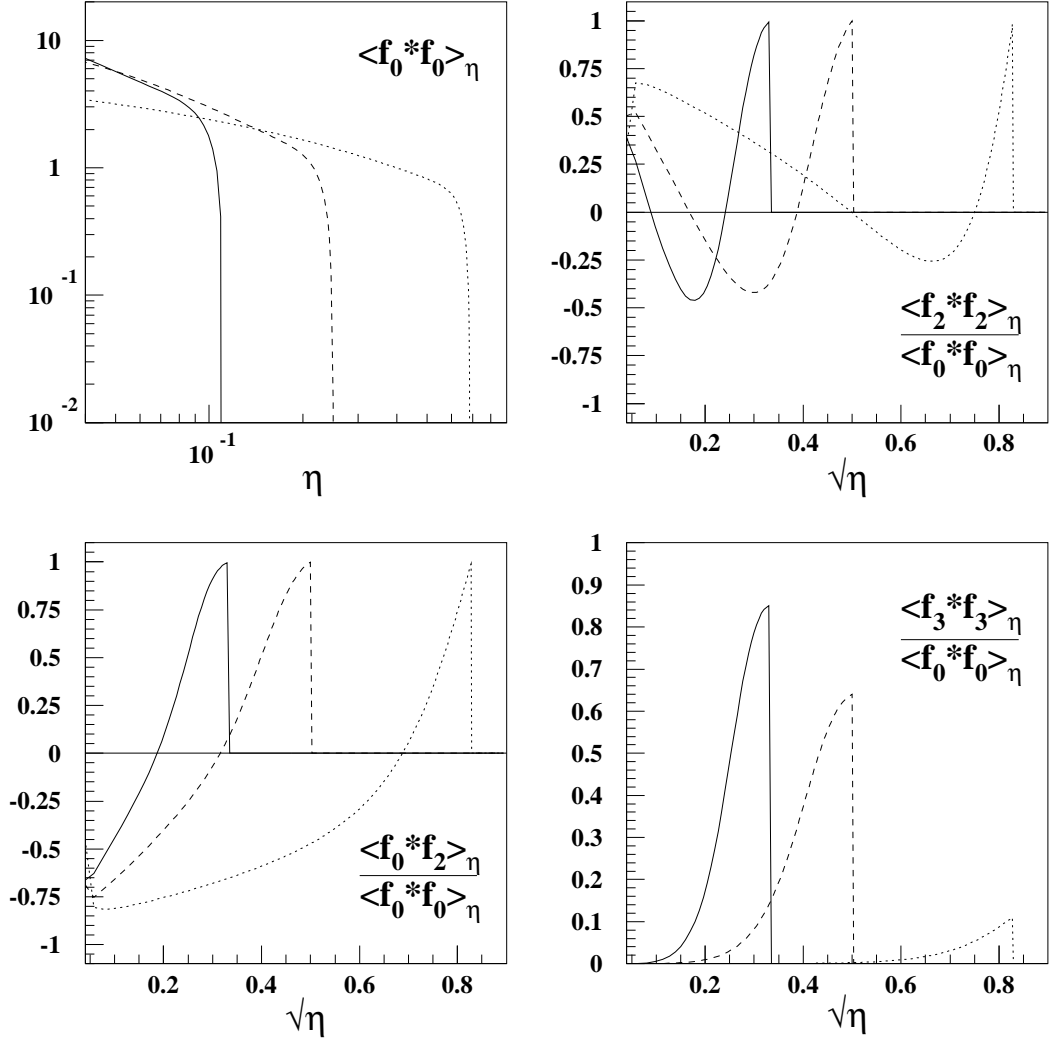,width=16cm,height=16cm}\hss}
\end{center}
\vskip -1.cm
\caption{ The luminosity correlation function $\langle f_0*f_0\rangle_\eta$ 
          and the three ratios of the correlation functions for three 
	  different values of $x$; $x=0.5$ (solid line), $1.0$ (dashed line), 
	  and $4.83$ (dotted line).}
\label{fig:fig1}
\end{figure}

\begin{figure}
\begin{center}
\hbox to\textwidth{\hss\epsfig{file=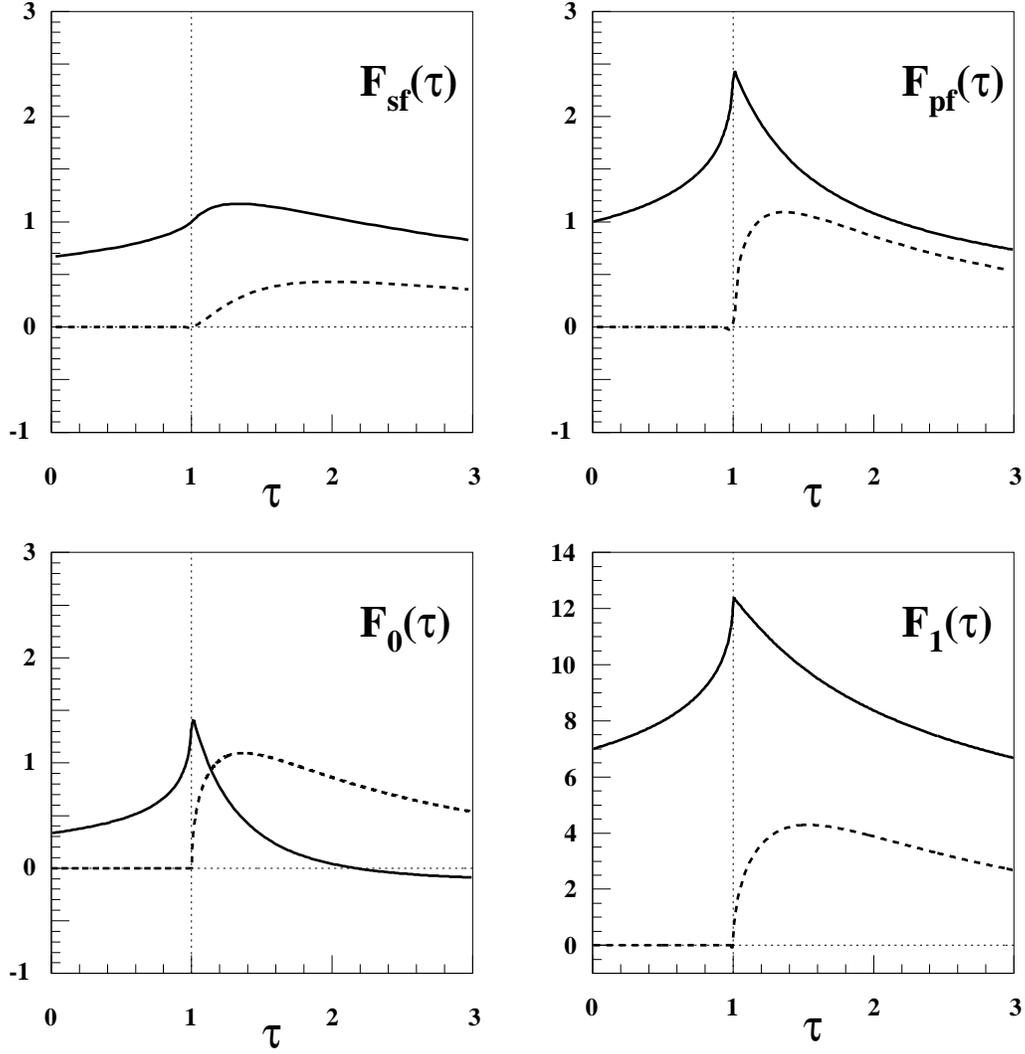,width=16cm,height=16cm}\hss}
\end{center}
\vskip -1.cm
\caption{The real (solid line) and imaginary (dotted line) parts of the 
         form factors $F_{sf}(\tau)$, $F_{pf}(\tau)$, $F_0(\tau)$ and
	 $F_1(\tau)$ as a function of the scaling variable $\tau$.}
\label{fig:fig2}
\end{figure}

\begin{figure}
\begin{center}
\hbox to\textwidth{\hss\epsfig{file=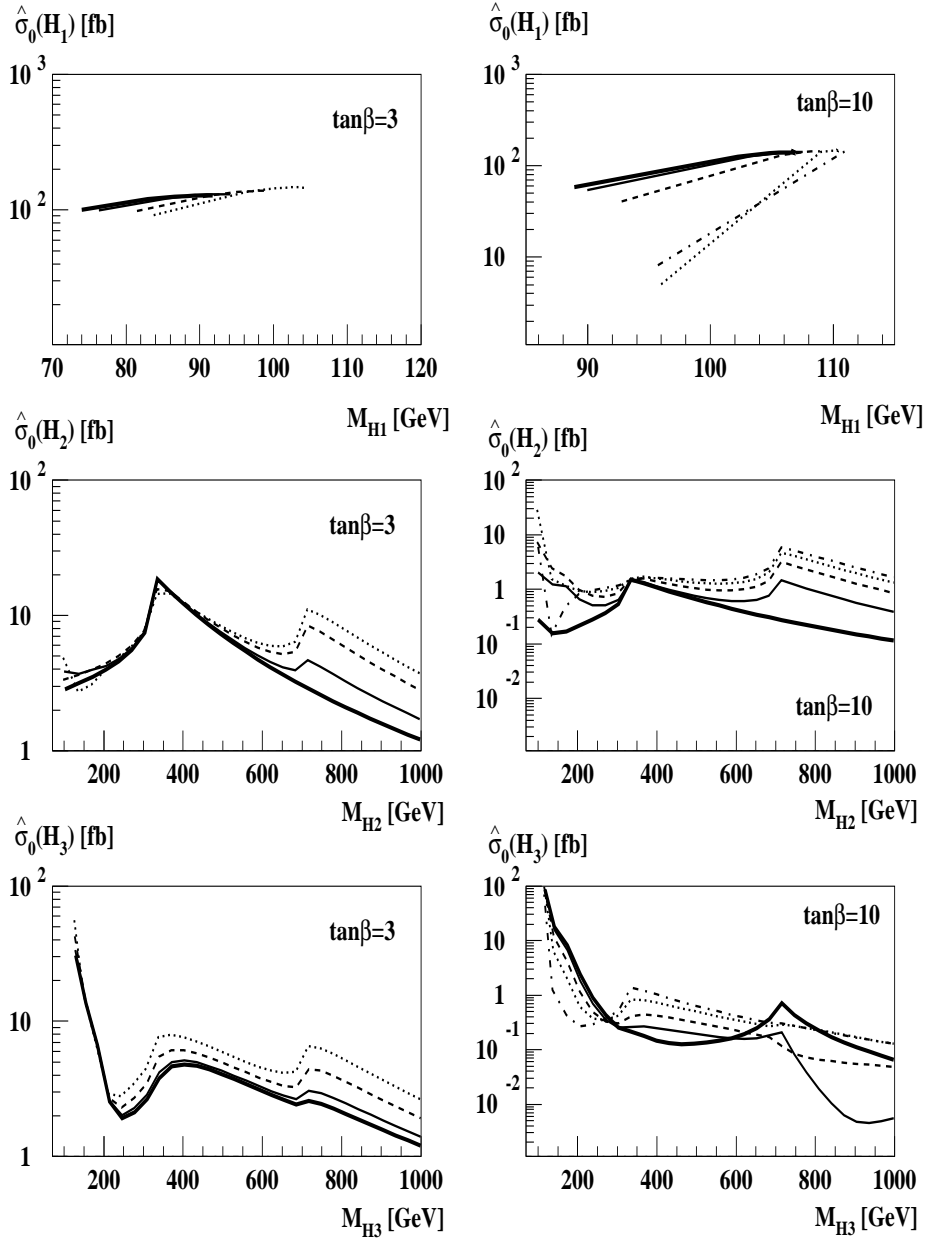,width=14cm,height=18cm}\hss}
\end{center}
\vskip -1.cm
\caption{The partonic cross sections $\hat\sigma_0(H_i)$ in units of fb 
         as a function of each Higgs--boson mass $M_{H_i}$ ($i=1,2,3$)
	 for five different values of the $CP$--violating phase;
	 $\Phi=0^{\rm o}$ (thick solid line), $40^{\rm o}$ (solid line), 
	 $80^{\rm o}$ (dashed line), $120^{\rm o}$ (dotted line), and 
	 $160^{\rm o}$ (dash--dotted line). The left column is for 
	 $\tan\beta=3$ and the right column for $\tan\beta=10$. }
\label{fig:fig3}
\end{figure}

\begin{figure}
\begin{center}
\hbox to\textwidth{\hss\epsfig{file=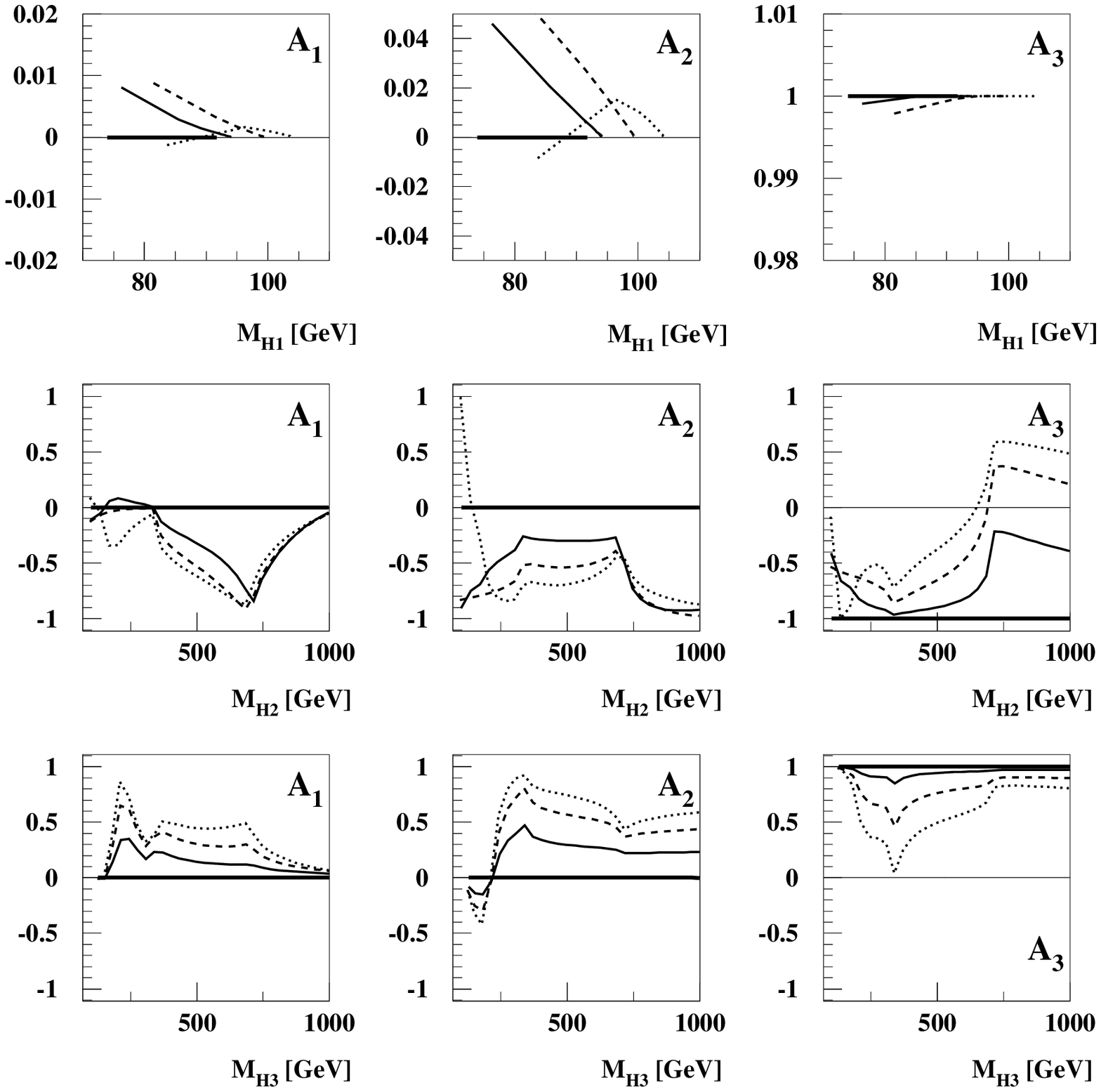,width=16cm,height=16cm}\hss}
\end{center}
\vskip -1.cm
\caption{The polarization asymmetries ${\cal A}_i$ as a function of
         each Higgs boson mass for four different values of the
	 $CP$--violating phase; $\Phi=0^{\rm o}$ (thick solid line),
         $40^{\rm o}$ (solid line), $80^{\rm o}$ (dashed line), 
	 and $120^{\rm o}$ (dotted line) in the case of 
         $\tan\beta=3$. The left column is for the asymmetry ${\cal A}_1$, 
	 the middle column for the asymmetry ${\cal A}_2$, and the right 
	 column for the asymmetry ${\cal A}_3$.}
\label{fig:fig4}
\end{figure}

\begin{figure}
\begin{center}
\hbox to\textwidth{\hss\epsfig{file=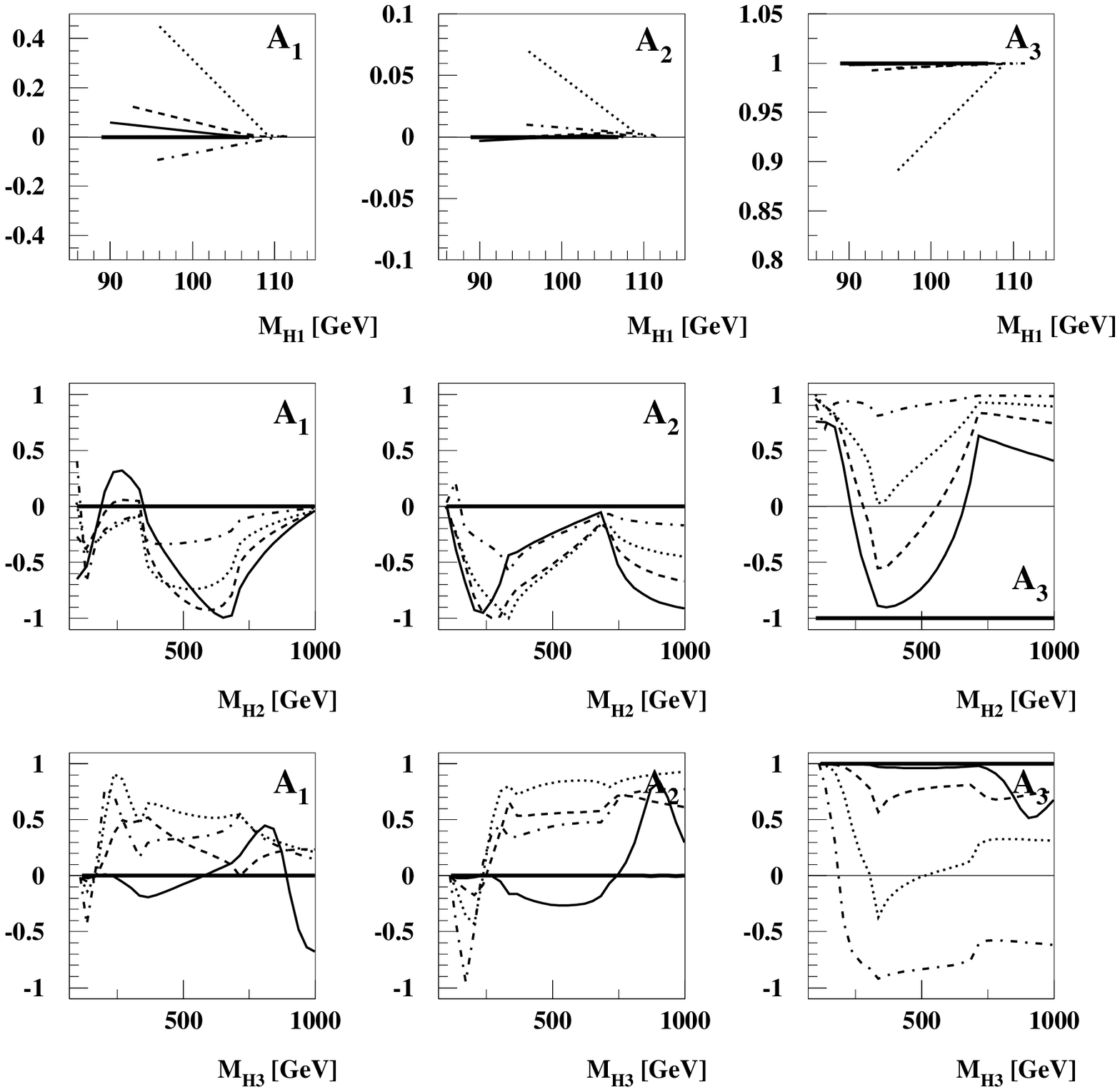,width=16cm,height=16cm}\hss}
\end{center}
\vskip -1.cm
\caption{The polarization asymmetries ${\cal A}_i$ as a function of
         each Higgs boson mass for five different values of the
	 $CP$--violating phase; $\Phi=0^{\rm o}$ (thick solid line),
         $40^{\rm o}$ (solid line), $80^{\rm o}$ (dashed line), 
	 $120^{\rm o}$ (dotted line), and $\Phi=160^{\rm o}$ (dash--dotted
	 line) in the case of for $\tan\beta=10$. 
	 The left column is for the asymmetry ${\cal A}_1$, 
	 the middle column for the asymmetry ${\cal A}_2$, and the right 
	 column for the asymmetry ${\cal A}_3$.}
\label{fig:fig5}
\end{figure}

%\end{multicols}


\begin{references}

\bibitem{DEMIR} D.A. Demir, Phys. Rev. D {\bf 60}, 095007 (1999);
   S.Y. Choi, hep--ph/9908397.

\bibitem{PW} A. Pilaftsis and C.E.M. Wagner, Nucl. Phys. {\bf B553}, 3 (1999);
   D.A. Demir, Phys. Rev. D {\bf 60}, 055006 (1999); B. Grzadkowski, 
   J.F. Gunion and J. Kalinowski, hep-ph/9902308.

\bibitem{IN} T. Ibrahim and P. Nath, Phys. Lett. B {\bf 418}, 98 (1998);
   Phys. Rev. D {\bf 57}, 478 (1998); D {\bf 58}, 019901 (1998) (E);
   {\it ibid}, 111301 (1998); W. Hollik, J.I. Illana, S. Rigolin,
   C. Schappacher, and D. Stockinger, Nucl. Phys. {\bf B439}, 3 (1999);
   T. Falk and K.A. Olive, Phys. Lett. B {\bf 439}, 71 (1998); 
   M. Brhlik, G.J. Good and G.L. Kane, {\it ibid.} 
   D {\bf 59}, 115004-1 (1999); S. Pokorski, J. Rosiek and C.A. Savoy, 
   hep--ph/9906206; A. Bartl, T. Gajdosik, W. Porod, P. Stochinger
   and H. Stremnitzer, hep-ph/9903402; E. Accomando, R. Arnowitt, and 
   B. Dutta, hep--ph/9907446; T. Ibrahim and P. Nath, hep--ph/9907555.

\bibitem{KAPLAN} S. Dimopoulos and G.F. Giudice, Phys. Lett. B {\bf 357}, 
   573 (1995); A. Cohen, D.B. Kaplan and A.E. Nelson, {\it ibid.} B {\bf 388},
   599 (1996); A. Pomarol and D. Tommasini, Nucl. Phys. {\bf B488}, 3 (1996);
   P. Bin\'{e}truy and E. Duda, Phys. Lett. B {\bf 389}, 503 (1996).

\bibitem{Many1} D.A. Demir, A. Masiero and O. Vives,
   Phys. Rev. Lett. {\bf 82}, 2447 (1999); Y.G. Kim, P. Ko and
   J.S. Lee, Nucl. Phys. {\bf B544}, 64 (1999); S. Baek and P. Ko, 
   Phys. Rev. Lett. {\bf 83}, 488 (1999); hep-ph/9904283; hep-ph/9907572; 
   A. Ali and D. London, hep-ph/9903535; hep-ph/9907243.
   
\bibitem{Many2} T. Falk, A. Frestl and K. Olive, hep-ph/9908311;
   T. Ibrahim and P. Nath, hep-ph/9907555; hep-ph/9908443;
   K. Freese and P. Gondolo, hep-ph/9908390.

\bibitem{Many3} J.F. Gunion, B. Grzadkowski, H.E. Haber and J. Kalinowski,
   Phys. Rev. Lett. {\bf 79}, 982 (1997); B. Grzadkowski, J. Gunion and 
   J. Kalinowski, Phys. Rev. D {\bf 60}, 075011-1 (1999); 
   S.Y. Choi, J.S. Shim, H.S. Song, and W.Y. Song,
   Phys. Lett. {\bf B449}, 207 (1999); S.Y. Choi, H.S. Song, and W.Y. Song, 
   hep--ph/9907474, to appear in Phys. Rev. D; S.Y. Choi, M. Guchait,
   H.S. Song and W.Y. Song, hep-ph/9904276; S.Y. Choi and J.S. Lee,
   hep-ph/9907496, to appear in Phys. Rev. D; 
   A. Pilaftsis, Phys. Rev. Lett. {\bf 77}, 4996 (1996); Nucl. Phys.
   {\bf B504}, 61 (1997); S.Y. Choi and
   M. Drees, Phys. Rev. Lett. {\bf 81}, 5509 (1998); S.Y. Choi and
   J.S. Lee, hep-ph/9909315.

\bibitem{TWOPHOTON} J.F. Gunion and H.E. Haber, Phys. Rev. D {\bf 48}, 
   5109 (1993); D.L. Borden, D.A. Bauer and D.O. Caldwell,
   {\it ibid.} D {\bf 48}, 4018 (1993);

\bibitem{GG} D. Grzadkowski and J.F. Gunion, Phys. Lett. B {\bf 294},
   361 (1992).

\bibitem{LINEAR} M. Kr\"{a}mer, J. K\"{u}hn, M.L. Stong and 
   P.M. Zerwas, Z. Phys. C {\bf 64}, 21 (1994);
   G.J. Gounaris and G.P. Tsirigoti, Phys. Rev. D {\bf 56}, 3030 (1997);
   D {\bf 58}, 059901 (1998) (E).

\bibitem{GKPST} I.F. Ginzburg, G.L. Kotkin, S.L. Panfil, V.G. Serbo, and
   V.I. Telnov, Nucl. Instr. and Math. {\bf 219}, 5 (1984).


\bibitem{CKP} D. Chang, W.-Y. Keung and A. Pilaftsis,
   Phys. Rev. Lett. {\bf 82}, 900 (1999); {\bf 83}, 3972 (1999) (E);
   A. Pilaftsis, hep-ph/9909485 v2; D. Chang, W.-F. Chang, and
   W.-Y. Keung, hep-ph/9910465.

\bibitem{HHG} See, for example, J.F. Gunion, H.E. Haber, G. Kane and S.
Dawson, {\sl The Higgs Hunters Guide} 
(Addison--Wesley Publishing Company, 1990).

\bibitem{SDGZ} M. Spira, A. Djouadi, D. Graudenz, and P.M. Zerwas,
  Nucl. Phys. {\bf B453}, 17 (1995) and references therein.

\bibitem{GLUE} S.Y. Choi and J.S. Lee, hep-ph/9910557.

\end{references}
\end{document}